\documentclass[useAMS,usenatbib,usegraphicx]{mn2e}
\newcommand{\Htwo}{H$_2$}
\newcommand\simlt{\lower.5ex\hbox{$\; \buildrel < \over \sim \;$}}
\newcommand\simgt{\lower.5ex\hbox{$\; \buildrel > \over \sim \;$}}

\title[An Observational Limit on the Earliest GRBs]{An Observational Limit on the Earliest GRBs}

\author[S. Naoz and O. Bromberg]{S. Naoz and O. Bromberg$^{1}$
\thanks{E-mail: smadar@wise.tau.ac.il (SN); omer@wise.tau.ac.il
(OB)}\\ $^{1}$School of Physics and Astronomy, The Raymond and Beverly
Sackler Faculty of Exact Sciences,\\ Tel Aviv University, Tel Aviv
69978, ISRAEL}

\begin{document}

\pagerange{\pageref{firstpage}--\pageref{lastpage}} \pubyear{2006}

\maketitle

\label{firstpage}

\begin{abstract}
We predict the redshift of the first observable (i.e., in our past
light cone) Gamma Ray Burst (GRB) and calculate the GRB-rate redshift
distribution of the Population III stars at very early times
($z=20-60$).  Using the last 2 years of data from \emph{Swift} we
place an upper limit on the efficiency ($\eta_{GRB}$) of GRB
production per solar mass from the first generation of stars.  We find
that the first observable GRB is most likely to have formed at
redshift 60. The observed rate of extremely high redshift GRBs (XRGs)
is a subset of a group of 15 long GRBs per year, with no associated
redshift and no optical afterglow counterparts, detected by
\emph{Swift}. Taking this maximal rate we get that
$\eta_{GRB}<1.1~10^{-4}$ GRBs per solar mass in stars. A more
realistic evaluation, e.g., taking a subgroup of $5\%$ of the total
sample of \emph{Swift} gives an upper limit of
$\eta_{GRB}<3.2~10^{-5}$ GRBs per solar mass.
\end{abstract}


\section{Introduction}\label{intro}

Gamma Ray Bursts (GRBs) are the brightest known events since the Big
Bang \citep[see reviews by:][]{piran05,piran00,rev}. Therefore they
offer a wonderful prospect to explore the evolution of the Universe
ever since stars began to form. The first generation of stars is
expected to be massive
\citep{Abel,BrCL02}, and at least some of them should form a GRB as
they end their lives \citep[e.g.,][]{Heger03}.  Since the luminosities
of galaxies and quasars decline with redshift, high redshift GRBs are
expected to be easier to observe than their host galaxies.
Consequently, they can offer  a significant probe of the early universe
\citep[e.g.,][]{Lamb00}. For example, since the GRB afterglows fade
only slowly with redshift they offer a source for studying the cosmic
reionization
\citep{BL04,BL06}. In particular, GRBs formation history 
is expected to follow the star formation evolution \citep{Blain,PM}.

The first stars in the Universe namely, population III (POP III),
formed at very high redshift  \citep[$z\simlt 66$,][]{NNB}. Since metals
are absent in the pre-stellar universe, the earliest available coolant
is molecular hydrogen (\Htwo). Thus the minimum halo mass that can
form a star is essentially set by requiring the infalling gas to reach
a temperature $>1000$~K required for exciting the rotational and
vibrational states of molecular hydrogen \citep{th2}. Numerical
simulations
\citep{Abel,fuller,Yoshida,reed,BrCL02} give a more accurate
constraint and require a minimum circular velocity $V_c\sim 4.5$~km/s,
where $V_c=\sqrt{GM/R}$ in terms of the halo virial radius $R$. These
simulations include gravity, hydrodynamic and chemical processes in
the primordial gas, and show that the first star formed within a
galactic halo of $\sim 10^5 M_{\odot}$ in total mass. The predicted
mass of this star is quite heavy and should exceed $100 M_\odot$
\citep{Yoshida06,Gao}.

The radiation from these first stars is
expected to eventually dissociate all the $H_2$ in the intergalactic
medium, leading to the domination of a second generation of larger
galaxies where the gas cools via radiative transitions in atomic
hydrogen and helium \citep{Haiman}. Atomic cooling occurs in halos
with $V_c > 16.5$ km/s, in which the infalling gas is heated above
10,000 K and is ionized.

\citet{BrL06} calculated the star formation evolution for population
III while assuming \emph{only} atomic cooling in order to derive the
GRB rate. They assumed that the efficiency of GRB production
($\eta_{GRB}$) is $\sim 10^{-9}$ GRBs per solar mass in stars, and
found that about $10\%$ of all bursts detected by
\emph{Swift} should be generated from redshift $\geq 5$. 
Unlike these writers and GRB rate analysis done by others
\citep[e.g.,][and references therein]{GPW05,GP05,GP07,BrL02,BrL06,DRM06}, 
we concentrate on a much higher redshift regime ($z=20-60$) and
we estimate the  upper bound of the efficiency 
of extremely
high redshift GRBs (XRGs) production. The article is structured as
follow: we begin by presenting a simple star formation evolution
(Section~\ref{sec:SFR}), relevant for these redshifts. Our analysis is
presented in Section~\ref{sec:pred}. In Section~\ref{sec:f_GRB} we
calculate the redshift of the first observable GRB.
Section~\ref{sec:eta} is dedicated to the calculation of the redshift
distribution of the XRG rate, in order to place an upper limit on
$\eta_{GRB}$.  We conclude with a discussion of our results
(Section~\ref{sec:diss}).

Our calculations are made in a $\Lambda$CDM universe, including dark
matter, baryons, radiation, and a cosmological constant. We assume
cosmological parameters matching the three year WMAP data together
with weak lensing observations \citep{Spergel06}, i.e.,
$\sigma_8=0.826$, $\Omega_m=0.299$, $\Omega_\Lambda=0.74$, and
$\Omega_b=0.0478$.

\section{Estimation of the High Redshift Star Formation Rate}\label{sec:SFR}

We adopt a simple model for the star formation history at high
redshift ($z\sim70-20$). This modle assumes that at these early times
most of the stars formed out of newly accreted gas during mergers. We
follow the linear and non linear perturbation growth from
\citet{NB05} and \citet{NNB} in order to derive the
fraction of mass in halos.

\citet{NB05} showed that the baryon sound speed varies
spatially, so that the baryon temperature and linear density
fluctuations must be tracked separately. In the non-linear regime, it
is necessary to calculate correctly the linearly extrapolated
overdensity $\delta_c$ which marks the time of the
collapse. \citet{NNB} showed that the value of $\delta_c$ is
lower at high redshift than the classical value ($1.686$) and varies
with time
\citep[see for reference fig.~6 in ][]{NB06}. Defining the mass variance
$S=\sigma^2(M,z)$, we calculate the \citet{Sheth} mass function, which
fits simulations and includes non-spherical effects on the
collapse. The function $f_{ST}$ is the fraction of mass associated
with halos of mass $M$:
\begin{equation}
\label{sheth}
f_{ST}(\delta_c,S)=A'\frac{\nu}{S}\sqrt{\frac{a'}{2\pi}}
\left[1+\frac{1}{\left(a'\nu^2\right)^{q'}} \right]
\exp \left[\frac{-a'\nu^2}{2} \right]\ ,
\end{equation}
where $\nu = \delta_c/\sqrt{S}$. We use best-fit parameters $a'= 0.75$
and $q'= 0.3$ \citep{Sheth02}, and ensure normalization to unity by
taking $A'= 0.322$.  We apply this formula with $\delta_c(z)$ and
$\sigma^2(M,z)$ as the arguments.

As mentioned, the only coolant available for these high redshift halos is
cooling via \Htwo~ cooling. Thus, we find the fraction of mass in halos
with mass larger than the minimum \Htwo~ cooling mass to be
\begin{equation}
F(>M_{min,H_2}(z))=\int{f_{ST}(\delta_c(z),S) dS}\ .
\end{equation}
Thus  the star
formation rate (SFR) (based on \Htwo~ cooling) is simply:
\begin{equation}
\label{sfr}
SFR=\frac{dF}{dt}\rho_0\frac{\Omega_b}{\Omega_m}\epsilon \ ,
\end{equation}
where $\rho_0$ is the comoving matter density, and $\epsilon$ is the
star formation efficiency. \citet{BL00} calculated the star formation
history for lower redshifts using a more complicated calculation
that included additional merger-induced star formation. \citet{BrL06}
have calculated separately the star formation rate for the different
star populations, i.e., POP I/II and POP III. In both papers the star
formation efficiency was chosen to be $10\%$ independent of
redshift. This yields a rough agreement between the SFR at low
redshift and the observations \citep[see fig.~1 in][]{BL00}. Following
these previous analyses we set $\epsilon=10\%$.

\begin{figure}
\centering
\includegraphics[width=84mm]{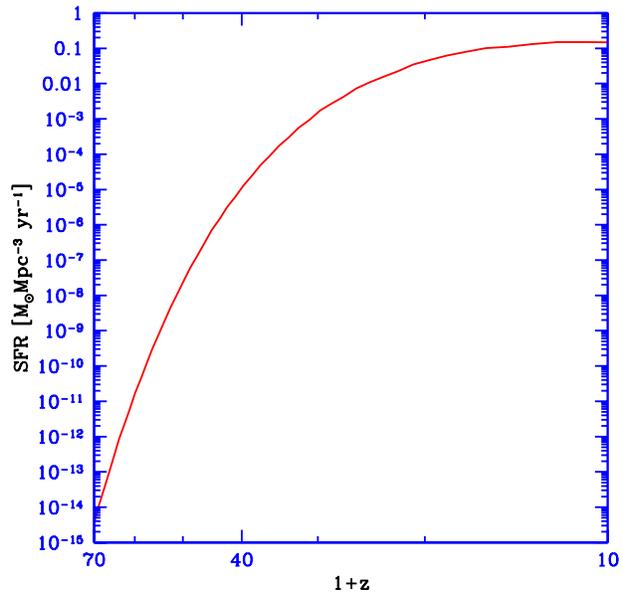}
\caption{Redshift evolution of SFR in halos larger than the minimum
\Htwo~ cooling mass, in $M_\odot$ yr$^{-1}$ per comoving Mpc$^3$.
This is the prediction from the model presented in equation
\ref{sfr}.}
\label{fig:SFR}
\end{figure}

In figure \ref{fig:SFR} we plot the SFR as a function of $1+z$ as
derived from eq.~\ref{sfr}. We note that this result is very close to
that obtained by
\citet{BrL02,BrL06} for the star formation of POP III. 
However our result predicts a higher rate because we consider star
formation at higher redshifts.  In practice our result is between
their most optimistic (high efficiency) estimate and their most
pessimistic case.  In addition our simple model does not include the
radiation feedback which lowers the contribution of molecular hydrogen
cooling compared to atomic cooling
\citep[see for reference][]{Haiman,BL01}. Therefore we restrict
ourselves to all redshifts $\geq 20$ (note that the redshift range
considered by \citet{BrL06} is $z\simlt 30$). Since we do not account
for reionization effects at all we compare our results to that of a
late reionization in \citep[figure 1 top panel in][]{BrL06}.

\section{Calculations and results}\label{sec:pred}

In order to produce a GRB the progenitor must fulfill 3 requirements
\citep[see ][for more detailed explanation]{z04,P05,BrL06}. First, the star
must be massive enough to result in a black hole. Second, the
star must lose its hydrogen envelope in order to allow the
relativistic jet to penetrate through the surface \citep{z04}, and
finally an accretion disk must be able to form. For that the star must
maintain enough angular momentum. When evaluating the efficiency of
producing GRBs from POP III stars there are many uncertainties, such
as the properties of the progenitor.
\citet{BrL06}, for example, claim that in order to produce a GRB, 
the progenitor must have a close binary companion that helps in 
stripping the hydrogen envelope and retaining sufficient angular 
momentum in the collapsing core. 
However, \citet{FWH01} show that massive POP III stars can produce
pair-instability supernovae that can strip off their hydrogen envelope
and result in a black hole accretion disk system. \citet{Heger03} deploy a
variety of masses that can produce GRBs from single POP III stars.
Whether or not binary stars are a strict requirement for GRB
production is still debatable. However, our analysis is independent of
that question. We only assume that GRBs in this high redshift regime
do exist, and use observational constrains to extract the
efficiency of producing them.

The most naive assumption is that each star (or  binary system) at
the relevant high redshift range produces a GRB and thus
$\eta_{GRB}\sim 1/100$ GRBs per solar mass (or $1/200$ assuming
binary progenitors). We derive further restrictions on $\eta_{GRB}$  
by normalizing the GRB rate with the potential XRG 
rate observed by \emph{Swift} (see text below for details).

\subsection{The First Observable GRB}\label{sec:f_GRB}

\citet{NNB} have found that the first star most likely  formed
at $z=65.8$.  As mentioned above, this star is expected to have a
mass $\simgt 100~M_\odot$ \citep{Abel}, thus it (as well as the stars
that follow it) is a suitable candidate to produce a GRB
\citep{Heger03}. We assume that GRBs at these redshifts are bright
enough to be observed today (estimates of the limiting luminosity are
given below using a more complex model), and adopt a fiducial jet
opening angle of $5^\circ$
\citep[e.g.,][]{Frl01}. This limited opening angle means that we see only
two out of $500$ GRBs. The probability to observe the first GRB on the
sky is computed following \citet{NNB} analysis for the formation of
the first star. Notice that since the lifetime of the first stars is
insignificant compared to the Hubble time, we can neglect the delay
time between the star formation and explosion.  We find that most
likely the first observable GRB is formed at $z=59.9$, with a
$1$-$\sigma$ ($68\%$) range of $z=58.9$--61.4 and a $2$-$\sigma$
($95\%$) range of $z=58.1$--63.5. Note that the most probable redshift
of the first GRB is only weakly sensitive to changing the opening
angle. For example an isotropic GRB will be seen at the redshift of
the first observable star. 
$20^\circ$ we find that
Considering an opening angle of $5^\circ$ we place an upper redshift
limit for observing GRBs. Combining with the limitation of our SFR
model we concentrate on the redshift range of: $20-60$.

\subsection{GRB production efficiency and rate}\label{sec:eta}

From equation (\ref{sfr}) we can evaluate the observable GRB rate:
\begin{equation}
\label{GRBR}
R_{GRB}(z)=\eta_{GRB}\frac{SFR}{1+z}\frac{dV}{dz}\int_{L_{min}(z)}^\infty
\phi(L)d\log L \ ,
\end{equation}
where $V$ is the comoving volume and the factor $(1+z)^{-1}$ accounts
for the cosmological time dilation. $\eta_{GRB}$ was defined earlier
as the GRB efficiency in units of GRBs per solar mass, and $\phi(L)$ is
the GRB luminosity function. The limiting luminosity at a given
redshift, $L_{min}$, depends on the flux sensitivity threshold of the
detector ($f_{lim}$), and can be written as:
\begin{equation}
\label{L_lim_1}
L_{min}=4\pi d_L^2(z)
f_{lim}\bigg\{\frac{C(E1(1+z),E2(1+z))}{C(E1,E2)}\bigg\}^{-1} \ ,
\end{equation}
where $d_L(z)$ is the luminosity distance and C(E1,E2) is the total
integrated luminosity for observation with \emph{Swift} between
$E1=50$ and $E2=300$ keV\footnote{The actual observed energy band of \emph{Swift}
is: $15-150$~keV, while our calculation follows
\citet{GP07}, who extrapolated to the BATSE energy  band.}
\citep[for specific explanation see for reference][]{GP07,sch99}. 

Assuming an average photon number  distribution of a
single power law, eq.~(\ref{L_lim_1}) can be written as \citep[see for
reference:][]{sch01}:
\begin{equation}
\label{L_lim_2}
L_{min}=4\pi d_L^2(z) f_{lim}(1+z)^{-(2+\bar{\alpha})} \ .
\end{equation}
where $\bar{\alpha}$ is the average photon spectral index over the
energy range (i.e., $dN/dE \propto E^{\bar{\alpha}}$, where $N$ is
the number of photons). A typical GRB photon number distribution can
be described as a broken power law, also known as a Band spectrum
\citep{Band93}. The low energy regime has a spectral slope of
$\alpha\simeq-0.8$, while the high energy regime has a slope of
$\beta\simeq-2.5$ \citep[][]{Preece00}.  The peak energy ($E_p$) falls
in the range of 100 keV to a few MeV, which renders it observable by
\emph{Swift}. 

Several groups \citep[e.g.,][]{Yon04,Ghir05,LDG05} using various
samples taken from BATSE, BeppoSax, Integral and HETE2, found that
$E_p\sim L_{iso}^{0.5}$, where $L_{iso}$ is the GRB isotropic
equivalent luminosity (either average or peak luminosity).  Each burst
that has an $E_p$ which falls within our observation window, can be
fitted with an average photon spectral slope,
$-0.8\simlt\bar{\alpha}\simlt-2.5$. Because $E_p$ is the break energy
in the power law and bursts with higher luminosity have higher $E_p$,
the average observed slope of such bursts should be higher (i.e.,
shallower) as well. \citet{sch01} found for a sample of simulated long
GRBs a median value of $\bar\alpha=-1.6$. In addition he found a
positive correlation between $\bar\alpha$ and the GRB luminosity, as
expected from the $E_p$-$L_{iso}$ relation. For example a typical XRG,
at $z=20$, with an intrinsic luminosity of $10^{53}~$erg~sec$^{-1}$
has an $E_p\sim1~$MeV which will be redshifted to $\sim50~$keV. The
photon spectrum within the BAT observational window will have an
average spectral slope of $\bar{\alpha}\simeq-1.6$.

We use this single power law to get a simplified k-correction for our
bursts, where we expect most XRG spectral slopes to fall between
$-1.6$ and $-1$, since their luminosity is in the high tail of the
luminosity function. We adopt a limiting case of $\bar\alpha=-1$,
according to the
\citet{sch01} hardness luminosity correlation\footnote{This is of course
the more interesting case since it favors higher luminosity bursts and
suits the redshift regime in this paper.}. We also address the
difference that adopting the  median value ($-1.6$) would make in our
results.

\citet{Lamb00} assumed that the \emph{Swift} threshold limit would be
$f_{lim}=0.04$ photons cm$^{-2}$~sec$^{-1}$, and that the spectral
photon index is $-1$. For an observed photon number flux with
an observed redshift they found the luminosity of the burst as well as
the maximum observable redshift. For example, for burst GRB971214 they
found a peak luminosity of $6.4\times 10^{58}$~sec$^{-1}$, and showed
that a burst with this luminosity can be seen up to redshift $\sim
70$. Repeating this exercise with the 2-year \emph{Swift} threshold limit
$f_{lim}=0.2$ photons cm$^{-2}$~sec$^{-1}$ we find that a burst with
this peak luminosity can be seen at least up to $z\sim 21$, while
adopting $\bar{\alpha}=-1.6$ yields a maximum observable
redshift of $7.8$. Thus, the maximum observed redshift for a given
luminosity is dependent on the spectral index.

\begin{figure}
\centering
\includegraphics[width=84mm]{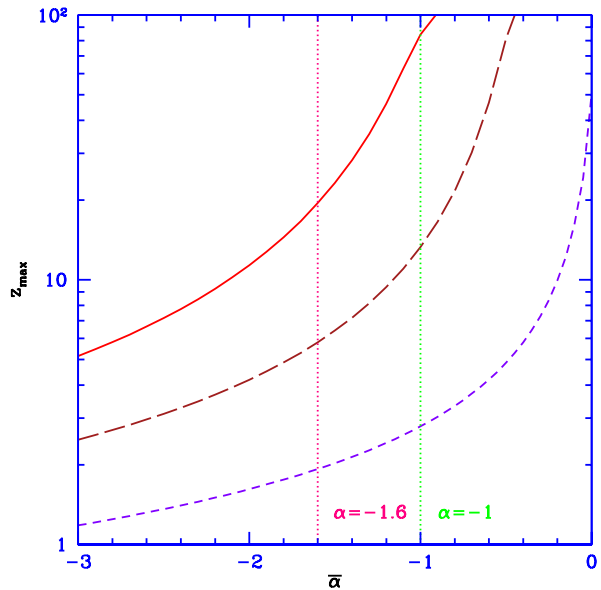} \caption{The maximum
detectable redshift by \emph{Swift} as a function of the spectral
photon index $\bar\alpha$, using the \emph{Swift} detection threshold
$f_{lim}=0.2$ photons cm$^{-2}$~sec$^{-1}$. We consider three
representative luminosities. From bottom to top: $10^{51},10^{52}$ and
$10^{53}$~ergs~sec$^{-1}$. Also shown are the median spectral index value from
\citet{sch01}, $\bar{\alpha}=-1.6$ and the limiting case
$\bar{\alpha}=-1$.}  \label{fig:max_z}
\end{figure}

In figure \ref{fig:max_z} we plot the maximum observable redshift of a
burst as a function of the spectral index, for three different
representative luminosities, from bottom to top: $10^{51},10^{52}$ and
$10^{53}$~ergs~sec$^{-1}$. The figure is plotted for $f_{lim}=0.2$
photons cm$^{-2}$~sec$^{-1}$, and it can be seen that the maximum
observable redshift of a fixed luminosity GRB is strongly dependent on
the spectral index.  Note that high luminosity bursts are observable
to higher redshift, especially since they tend to have high
$\bar{\alpha}$. In the figure we indicate the limiting value
and the median value of $\bar{\alpha}$ , $-1$ and $-1.6$,
respectively. Thus, high luminosity bursts ($\sim
10^{53}$~ergs~sec$^{-1}$) with $\bar{\alpha}\sim -1$ can be expected
to be observed up to $z\sim 80$, while for the median value of the
spectral index $-1.6$ the bursts are expected to be observed up to
$z\sim 18$. Thus observing GRBs from a higher redshift (while assuming
$\bar{\alpha}=-1.6$) suggests much more luminous bursts than $\sim
10^{53}$~ergs~sec$^{-1}$. Of course this can also mean (when placing
an upper limit on the luminosity) that there are no observable
extremely high redshift GRBs with the current
\emph{Swift} detection limit. Therefore for \emph{Swift} to detect
XRGs, their average spectral slope should be closer to $-1$, as
shown in fig. \ref{fig:max_z}.

We assume the \citet{GPW05} and \citet{GP07} luminosity function based
on \citet{sch01}.  This luminosity function has a broken power law with
cutoff luminosities of $L^\star/\Delta_1$ and $\Delta_2L^\star$.
\begin{equation}
\label{Phi_LG}
\phi(L)=c_0\left\{ \begin{array}{ll}
\left(\frac{L}{L^\star} \right)^a & \frac{L^\star}{\Delta_1}<L<L^\star \ , \\
\left(\frac{L}{L^\star} \right)^b & L^\star<L<\Delta_2 L^\star \ ,\\
\end{array}\right.
\end{equation}
where $\phi(L)$ is normalized so that the integral over the luminosity
function equals unity through $c_0$. We have used the \citet{GP07} best
fitted values with $\Delta_1,\Delta_2=100$, $L^\star=6.5\times
10^{51}$~erg~sec$^{-1}$, $a=-0.2$ and $b=-1.7$, marked in
their paper as model (v). This model favors high luminosity bursts,
which are the interesting case here. We also address their (vi) model
with $b=-2$ which has fewer high luminosity bursts (see text
below).

We can now return to eq.~(\ref{GRBR}) and calculate the XRG rate
from POP III stars at $z=60-20$, and use that to place an upper
limit of the efficiency of GRB production from these stars.  We ask
what is the efficiency of GRB production assuming that 
\emph{Swift} observes up to $\sim 15$
\emph{long} GRBs per year originating from these high redshift POP III
stars. This subsample includes GRBs with no associated redshift (there
are about 60 per year out of 90 per year). In addition we exclude GRBs
with $T_{90}$ shorter than $50~$sec\footnote{A burst with $T_{90}=50$
at $z=20$ has a proper time duration of $\sim 2.5~$sec which places it
on the border between long and short GRBs.}, where $T_{90}$ defines
the duration of the burst, since short GRBs most likely originate from
much lower redshift
\citep[see for reference:][]{udi}.
Finally we exclude GRBs with no optical counterparts for the following
reason: consider a burst at $z\sim20$, with an afterglow counterpart.
The photons of the burst redshift with the expansion of the Universe
as they propagate. Wavelengths shorter than the Lyman limit will be
absorbed by photoionizing hydrogen or helium atoms. Assuming that the
Universe was completely reionized at $z\sim 10$ or later each $z=0$
optical photon was at a wavelength shorter than the Lyman limit and
thus was absorbed in the pre-reionization era \citep[see for
reference:][]{BL01}.  Therefore, no optical afterglow should be
observed from XRGs, and we are left with about 15 bursts per year from
the \emph{Swift} sample.  Of course assuming that all of this sample
originates from high redshift is mostlikely an
overestimate. Nevertheless, it gives an upper limit to the
efficiency. Below we give more realistic estimates of the XRG rate
detected by \emph{Swift}.

\begin{figure}
\centering
\includegraphics[width=84mm]{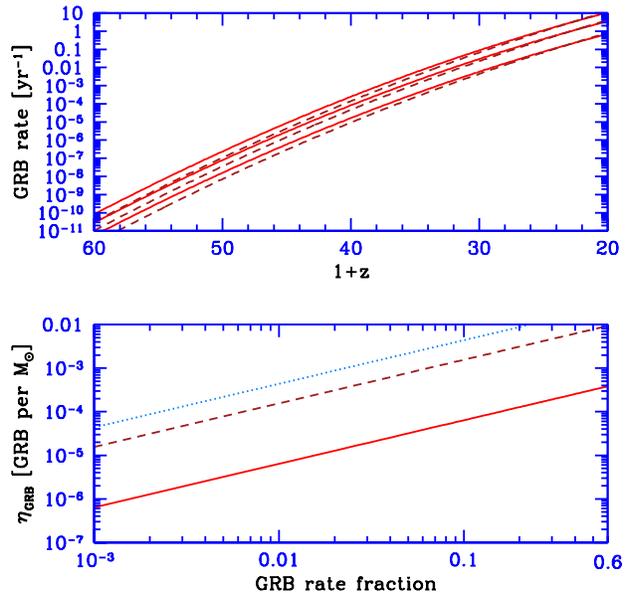}
\caption{Top panel: Redshift distribution of GRBs for various
POP III contribution to the total \emph{Swift} detection sample. We consider
a spectral index of $-1$ (solid curves) and also a spectral index of
$-1.6$ (dashed curves). For the first case ($\bar{\alpha}=-1$) we
consider a contribution of (top to bottom) $15,5$ and $1$ GRBs per
year from redshifts of $20-60$, which results in
$\eta_{GRB}=1.1~10^{-4}$,$3.5~10^{-5}$ and $7.1~10^{-6}$ GRBs per solar
mass respectively. For the second case ($\alpha=-1.6$) we consider
(top to bottom) $15,5$ and $1$ GRBs per year from the relevant
redshifts. The resulting upper limit on the efficiency is
$\eta_{GRB}=2.6~10^{-3},8.7~10^{-4}$ and $1.7~10^{-4}$ GRBs per solar
mass, respectively.  Lower panel: $\eta_{GRB}$ for various fractional
 contribution of GRBs out of $\sim 90$ GRBs per year detected by
\emph{Swift}, from redshifts $20-60$.  We consider the spectral index
of $-1$ (solid curve) and also spectral index of $-1.6$ (dashed
curve). For this latter case we also consider the parameters of model
(vi) in \citet{GP07} (dotted curve), which is less favorable for high
luminosity bursts ($b=-2$). }
\label{fig:GRBr}
\end{figure}

In figure
\ref{fig:GRBr} top panel, we depicted the predicted GRB rate redshift 
distribution for various efficiency normalizations. We consider both
cases of $\bar{\alpha}=-1$ (solid curves) as well as
$\bar{\alpha}=-1.6$ (dashed curves). For the first case we consider
XRGs that contribute  (from top to bottom)
$15,5$ and $1$ GRBs per year to the total sample detected by
\emph{Swift}. The GRB efficiencies from these values are:
$\eta_{GRB}=1.1~10^{-4}$,$3.5~10^{-5}$ and $7.1~10^{-6}$ GRBs per
solar mass, respectively.  For the latter case ($\bar{\alpha}=-1.6$) we
consider the same contributions, (top to bottom) $15,5$ and $1$ GRBs
per year from the relevant redshift. The resulting upper limit on the
efficiency is $\eta_{GRB}=2.6~10^{-3},8.7~10^{-4}$ and $1.7~10^{-4}$
GRBs per solar mass, respectively. Not presented in the figure is the
lower limit case of $\bar{\alpha}=-2.2$. We find that for this
spectral index
\emph{Swift} will be able to detect only a few GRBs per year. Comparing
with the results above, in this case the efficiency needed to
contribute one GRBs per year detectable by \emph{Swift} is
$4.2~10^{-3}$ GRBs per solar mass.

As can be seen from the figure in order to generate the same
contribution of GRBs per year, the efficiency must be different for
different spectral indices. Adopting $\bar{\alpha}=-1$ generates a
lower efficiency for the same percentage of GRB contribution. This can
be seen more clearly in the bottom panel of this figure, where we
depict the GRB efficiency per solar mass ($\eta_{GRB}$) for various
fractional contribution of GRBs from redshifts $20-60$ (out of $\sim
90$ GRBs per year detected by \emph{Swift}). We consider the spectral
index of $-1$ (solid curve) and also spectral index of $-1.6$ (dashed
curve). For example, for a $1\%$ contribution of GRBs detected by
\emph{Swift} per year from this redshift range, i.e., about one GRB
per year, we find that a spectral index of $-1$ predicts about $2$
orders of magnitude lower efficiency of GRB production than assuming a
spectral index of $-1.6$. As mentioned before, the latter spectral index
suggests much more luminous GRBs which is less reasonable.  We also
consider (in this bottom panel of figure \ref{fig:GRBr}) the
parameters of model (vi) in
\citet{GP07} (doted curve), which is less favorable for high
luminosity bursts ($b=-2$). As can be expected this model results in a
higher GRB efficiency since the luminosity function does not favor
high luminosity bursts.

\citet{BrL06} found that about $10 \%$ of the GRBs detected by 
\emph{Swift} each 
year should result from high redshift ($\geq 5$) stars. However they
assumed a lognormal luminosity function, and efficiency of $\sim
10^{-9}$ GRBs per solar mass. Thus if this $10 \%$ claim is true, it
is reasonable to assume that even less originate from a higher
redshift ($\geq 20$) range. Suppose we assume a more reasonable upper
limit for example, about $5
\%$ to the total sample (about half of their  range) .  Using their luminosity
function \citep[see eq.~8 in][]{BrL02} we find that for
$\bar{\alpha}=-1$, this $5\%$ contribution (about five GRBs per year)
results in an upper limit on the efficiency of: $1.8~10^{-6}$ GRBs per
solar mass, while assuming $\bar{\alpha}=-1.6$ gives an efficiency of
$1.0~10^{-5}$ GRBs per solar mass. Adopting the
\citet{GPW05} and \citet{GP07} luminosity function, (eq.~(\ref{Phi_LG}))
we have (as depicted in fig.~\ref{fig:GRBr})
$\eta_{GRB}=3.2~10^{-5}$ and $7.8~10^{-4}$ GRBs per solar mass, for
$\bar{\alpha}=-1$ and $-1.6$, respectively.

\section{DISCUSSION}\label{sec:diss}

We have predicted that the redshift of the first observable GRB is most
likely $z\sim 60$. This prediction is based on the \citet{NNB}
prediction of the redshift of the first observable star.

We have calculated the GRB rate from extremely high redshift
($z=20-60$) POP III stars. We used a simple SFR (based on
\Htwo~cooling) that includes the correct evolution of fluctuations.
We assume that at these early times most of the stars formed out of
newly accreted gas during mergers. We find that our SFR is consistent
with previous analysis at lower redshift \citep{BrL06}. However our
result predicts a higher rate (see fig.~\ref{fig:SFR}) because we
consider star formation at higher redshifts.

Using this SFR we evaluated the GRB rate redshift distribution
(eq.~(\ref{GRBR})).  We adopted \citet{GP05} luminosity function, and
performed a simple k-correction.  We showed that the maximum redshift
to which GRBs can be observed depends on their luminosity and on the
spectral index (see fig.~\ref{fig:max_z}). Using also the
$E_p$-$L_{iso}$ relation and, as supported by the \citet{sch01}
luminosity hardness correlation, we adopted $\bar{\alpha}=-1$. This
value enables high luminosity bursts to be seen to larger distances,
i.e., higher redshift. 

Using the \emph{Swift} detection rate we have placed an upper limit on
the efficiency of GRB production from POP III stars. For various
fractional contribution to the total \emph{Swift} sample we find different
efficiencies (fig.~\ref{fig:GRBr}). The safest upper limit we set is:
$1.1~10^{-4}$ GRBs per solar mass. This is a result from a maximal
contribution of XRGs to the \emph{Swift} sample. This sub-sample has
$15$ very long ($T_{90}>50$~sec) GRBs per year with no associated
redshift and no optical afterglow counterparts (about $\sim 17\%$, of
the total \emph{Swift} detection sample).  We also find a more
realistic estimate, that the upper limit of GRB efficiency from POP
III stars is $3.2~10^{-5}$ GRBs per solar mass. This efficiency
results in a high redshift GRB contribution of $5\%$ of the
\emph{Swift} sample. Taking the median value $\bar\alpha=-1.6$ we
find that the upper limit on $\eta_{GRB}$ is $7.8~10^{-4}$ GRBs per
solar mass.

We caution that we have extrapolated the properties of the observed
GRBs to much higher redshift. An additional, important caveat to our
model is the strong dependence on the spectral index
$\bar{\alpha}$. We point out that given the $E_p$-$L_{iso}$ relation
it is reasonable to assume that $\bar\alpha\simgt -1.6$. The
$E_p$-$L_{iso}$ relation is closly related to the \citet{Am02}
relation. Several authors
\citep[e.g.,][]{BP05,NP05} claimed that this relation may be a
boundary curve created by the brightest and softest bursts, as a
result of observational biases. They suggested that many bursts should
be dimmer and harder \citep[though see][]{Am06}. This supports our
argument that XRGs should have high $\bar\alpha$.  For completeness,
though, we have shown that if $\bar{\alpha}=-2.2$,
\emph{Swift} can detect only a few GRBs per year from redshift 
$>20$, and for \emph{one} GRB per year we find a high upper limit on
the efficiency of $4.2~10^{-3}$ GRBs per solar mass.

\section*{Acknowledgments}
The authors would like to offer special thanks to Rennan Barkana, Avi
Loeb, Ehud Nakar and Eran Ofek for meaningful discussions. SN
acknowledges support by Israel Science Foundation grant 629/05 and
U.S. - Israel Binational Science Foundation grant 2004386, and OB
thanks the ISF grant for the Israeli Center for High Energy
Astrophysics.

\bsp

\label{lastpage}

\end{document}